\documentclass{aastex63}
\usepackage{CJK}
\usepackage{amsmath}
\usepackage{multirow}

\received{...}
\revised{...}
\accepted{...}

\shorttitle{NS Mass Function}
\shortauthors{Li et al.}

\begin{document}
\begin{CJK*}{UTF8}{gbsn}

\title{Population properties of neutron stars in the coalescing compact binaries}

\author[0000-0001-5087-9613]{Yin-Jie Li （李银杰）}
\affiliation{Key Laboratory of Dark Matter and Space Astronomy, Purple Mountain Observatory, Chinese Academy of Sciences, Nanjing 210023, People's Republic of China}
\affiliation{School of Astronomy and Space Science, University of Science and Technology of China, Hefei, Anhui 230026, People's Republic of China}

\author[0000-0001-9120-7733]{Shao-Peng Tang（唐少鹏）}
\affiliation{Key Laboratory of Dark Matter and Space Astronomy, Purple Mountain Observatory, Chinese Academy of Sciences, Nanjing 210023, People's Republic of China}
\affiliation{School of Astronomy and Space Science, University of Science and Technology of China, Hefei, Anhui 230026, People's Republic of China}

\author[0000-0001-9626-9319]{Yuan-Zhu Wang（王远瞩）}
\affiliation{Key Laboratory of Dark Matter and Space Astronomy, Purple Mountain Observatory, Chinese Academy of Sciences, Nanjing 210023, People's Republic of China}

\author[0000-0001-9034-0866]{Ming-Zhe Han  (韩明哲)  }
\affiliation{Key Laboratory of Dark Matter and Space Astronomy, Purple Mountain Observatory, Chinese Academy of Sciences, Nanjing 210023, People's Republic of China}
\affiliation{School of Astronomy and Space Science, University of Science and Technology of China, Hefei, Anhui 230026, People's Republic of China}

\author[0000-0003-4891-3186]{Qiang Yuan（袁强）}
\affiliation{Key Laboratory of Dark Matter and Space Astronomy, Purple Mountain Observatory, Chinese Academy of Sciences, Nanjing 210023, People's Republic of China}
\affiliation{School of Astronomy and Space Science, University of Science and Technology of China, Hefei, Anhui 230026, People's Republic of China}

\author[0000-0002-8966-6911]{Yi-Zhong Fan（范一中）}
\affiliation{Key Laboratory of Dark Matter and Space Astronomy, Purple Mountain Observatory, Chinese Academy of Sciences, Nanjing 210023, People's Republic of China}
\affiliation{School of Astronomy and Space Science, University of Science and Technology of China, Hefei, Anhui 230026, People's Republic of China}
\email{The corresponding author: yzfan@pmo.ac.cn (Y.Z.F)}

\author[0000-0002-9758-5476]{Da-Ming Wei（韦大明）}
\affiliation{Key Laboratory of Dark Matter and Space Astronomy, Purple Mountain Observatory, Chinese Academy of Sciences, Nanjing 210023, People's Republic of China}
\affiliation{School of Astronomy and Space Science, University of Science and Technology of China, Hefei, Anhui 230026, People's Republic of China}

\begin{abstract}
We perform a hierarchical Bayesian inference to investigate the population properties of the coalescing compact binaries involving at least one neutron star (NS). 
With the current gravitational wave (GW) observation data, we can rule out none of the Double Gaussian, Single Gaussian, and Uniform NS mass distribution models, though a specific Double Gaussian model inferred from the Galactic NSs is found to be slightly more preferred.
The mass distribution of black holes (BHs)  in the neutron star-black hole (NSBH) population is found to be similar to that in the Galactic X-ray binaries. 
Additionally, the ratio of the merger rate densities between NSBHs and BNSs is estimated to be $\sim 3:7$.
The spin properties of the binaries, though constrained relatively poorly, play nontrivial role in reconstructing the mass distribution of NSs and BHs. We find that a perfectly aligned spin distribution can be ruled out, while a purely isotropic distribution of spin orientation is still allowed. To evaluate the feasibility of reliably determining the population properties of NSs  in the coalescing compact binaries with upcoming GW observations, we perform simulations with a mock population. We find that with 100 detections (including BNSs and NSBHs) the mass distribution of NSs can be well determined, and the fraction of BNSs can also be accurately estimated.
 
\end{abstract}

\keywords{Gravitational waves---Neutron stars---Binaries: close}

\section{Introduction} \label{sec:intro}
Recently, the observations of gravitational wave (GW) signals from two neutron star-black hole (NSBH) coalescences, i.e., GW200105 and GW200115, have been reported by LIGO/Virgo/KAGRA Collaboration \citep[LVKC;][]{2021ApJ...915L...5A}. These two events are also the first confident observations of NSBH binaries in the Universe, although previously the modeling of the kilonova emission of hybrid GRB 060614  was also strongly in favor of an NSBH merger origin \citep{2015NatCo...6.7323Y,2015ApJ...811L..22J}. Since the first successful detection of GW \citep{2016PhRvL.116f1102A} in 2015, about $\sim 50$ compact binary coalescences (CBCs) have been formally reported \citep{2021PhRvX..11b1053A}, including $\sim 45$ events from binary black hole (BBH) coalescences and several mergers involving at least one neutron star \citep[e.g.,][]{2017PhRvL.119p1101A,2020ApJ...892L...3A,2020ApJ...896L..44A,2020PhRvL.125j1102A}. Very recently, \cite{2021arXiv210801045T} releases a deep extended catalog, which reports another 8 new events. With a rapidly increasing sample of GW events, our understanding of the population properties of the stellar BHs in the Universe has been advanced \citep[e.g.,][]{2017PhRvD..96b3012T,2019ApJ...882L..24A,2020ApJ...891L..27F,2021ApJ...913L...7A,2021ApJ...913...42W,Li_2021,2021ApJ...915L..35K,2021ApJ...913L..19T,2021ApJ...915L..35K,2021arXiv210902424G}, and some formation/evolution processes of the compact binaries are being revealed \citep[e.g.,][]{2020ApJ...900..177K,2021ApJ...915L..35K,2021ApJ...907L..24S,2021ApJ...916L..16B,2021arXiv210708811T,2021arXiv210906222M,2021arXiv211010838W}.
However, due to the limited observation of NSBH and BNS mergers up to now, the population properties of these kinds of coalescence systems are hard to probe. Moreover, \cite{2020ApJ...892...56T} found that the misclassification of BBH into NSBH makes it more challenging to reconstruct the mass function of the BHs. 

Thanks to the observation of Galactic radio pulsars, a large number of NS masses have been measured \citep{2011MNRAS.414.1427V,2016arXiv160501665A,2016ARA&A..54..401O,2018MNRAS.478.1377A,2019ApJ...876...18F,2019AN....340..957R,2020PhRvD.102f3006S,2021ApJ...909L..19G,2020RNAAS...4...65F}. The mass distribution of these Galactic NSs has been investigated and  a double gaussian model with two peaks at $\sim1.35M_{\odot}$ and $\sim1.9M_{\odot}$ are found to be able to well reproduce the data  \citep{2018MNRAS.478.1377A,2020PhRvD.102f3006S}. Nevertheless, it is very interesting to investigate the mass distribution of NSs in the Universe, and the GW observations provide the unprecedented valuable chance. Recently, \cite{2021arXiv210704559L} tried to reconstruct the mass distribution of this extragalactic population of NSs and found that the mass function seems to be more consistent with a uniform distribution than the bimodal Galactic population. In their hierarchical inferences, only the masses of compact objects have been taken into account,  and the spin information of the compact binaries has not been included. However, as found in the literatures \citep[e.g.][]{2013PhRvD..87b4035B,2018JCAP...03..007C,2020PhRvR...2d3096P}, for the GW events involving at least one BH, there is mass-spin degeneracy.
Additionally, the measurements of both the component masses and misaligned spins are of prime importance in understanding the origin and evolution of astrophysical compact binaries \citep{2016ApJ...832L...2R,2018A&A...616A..28Q,2019ApJ...870L..18Q}. 
Binaries born in isolated evolution are expected to form with nearly aligned spins \citep{2000ApJ...541..319K}, and after considering the supernova kicks, the compact object binaries tend to still have small misalignments \citep{2016ApJ...832L...2R,2018PhRvD..98h4036G,PhysRevD.97.043014}. On the contrary, binaries originating from dynamical capture are expected to have isotropic distribution for their spin orientations \citep{2017CQGra..34cLT01V,2016ApJ...832L...2R}. By surveying the spin properties of the BBH population, \cite{2021ApJ...913L...7A} find that the current BBH merger events likely originate from both formation channels \citep[see also][]{2021PhRvD.103h3021W,2021ApJ...910..152Z}.

In this work, we perform a hierarchical Bayesian inference to investigate the population properties of NSs detected by LIGO and Virgo with the information of their masses and spins. In Sec. \ref{sec:method}, we introduce the data and the models used for inference, and in Sec. \ref{sec:result} we present the results. In Sec. \ref{sim}, we carry out simulations with a mock population and evaluate the feasibility for reconstruction of the NS mass distribution with one hundred of  BNSs/NSBHs detected in the design sensitivity run of Advanced LIGO/Virgo, and Sec. \ref{sec:discussion} is our conclusion and discussion.   

\section{Method} \label{sec:method}

\subsection{Selected events}\label{method_data}
Our hierarchical analysis is based on the GW events involving at least one NS, which consist of four published confident events, including GW170817, GW190425, GW200105 and GW200115  \citep{2017PhRvL.119p1101A,2020ApJ...892L...3A,2021ApJ...915L...5A}, and one marginal event, GW190426\_152155. Though the nature of GW190425 and GW190426\_152155 are not fully solved yet \citep{2020ApJ...891L...5H,2020arXiv201204978L}, the former (latter) is more likely a BNS (NSBH) \citep{2020ApJ...892L...3A,2021PhRvX..11b1053A}. We do not include GW190814 in our analysis, because the NS nature of the secondary object in GW190814 is inconsistent with either the maximum mass of nonrotating NS \citep{2021ApJ...908L..28N,2021PhRvD.104f3032T}, determined by the multi-messenger analyses of GW170817/GRB 170817A/AT2017gfo \citep{2018PhRvD..97b1501R,2018ApJ...852L..25R,2019PhRvD.100b3015S,2020PhRvD.101f3029S,2020ApJ...904..119F} or the constraints obtained from energetic heavy-ion collisions \citep{2020PhRvC.102f5805F}. Though a rapidly rotating NS may be allowed to have the mass of the secondary object in GW190814, it will coalesce to a BH before merger due to the rotational instabilities \citep{2020PhRvD.101f3029S,2021MNRAS.505.1600B}.
Therefore, our data set includes two BNS events and three NSBH events (or candidate), and the parameter estimation results for each event are adopted from \cite{2019PhRvX...9c1040A,2021PhRvX..11b1053A} \footnote{Download from \url{https://dcc.ligo.org/LIGO-P2000223/public}, \url{https://dcc.ligo.org/LIGO-P1800370/public}, and \url{https://dcc.ligo.org/LIGO-P2100143/public}}. Since we need the spin information (including spin magnitude and orientation) for each compact object, all the posterior samples should be estimated by the precession waveforms. So in this work we adopt the ``IMRPhenomPv2NRT\_lowSpin\_posterior", ``PhenomPNRT-LS", ``C01:PhenomXPHM\_low\_spin", ``C01:PhenomXPHM\_low\_spin", and ``PrecessingSpinIMRHM" posterior samples for GW170817, GW190425, GW200105, GW200115, and GW190426\_152155, respectively.

\subsection{Models} \label{method_model}
For the hierarchical population inference, here we consider three typical models for NS mass distribution and two for BH mass distribution. The first NS mass function model is the Double Gaussian scenario, 
\begin{equation}\label{DG}
\pi(m_{\rm NS}|m_{\rm min}, m_{\rm max}, \mu_1, \sigma_1, \mu_2, \sigma_2, r_1) = r_1\mathcal{N}_1(m_{\rm NS}|\mu_1,\sigma_1)/\Phi_1 + (1-r_1)\mathcal{N}_2(m_{\rm NS}|\mu_2,\sigma_2)/\Phi_2, ~ \text{for}~ m_{\rm NS} \in{(m_{\rm min}, m_{\rm max})} .
\end{equation} 
The second is the Single Gaussian model,
\begin{equation}
\pi(m_{\rm NS}|m_{\rm min}, m_{\rm max}, \mu, \sigma) = \mathcal{N}(m_{\rm NS}|\mu,\sigma)/\Phi, ~ \text{for}~ m_{\rm NS} \in{(m_{\rm min}, m_{\rm max})} .
\end{equation} 
And the third is the Uniform scenario,
\begin{equation}
\pi(m_{\rm NS}|m_{\rm min}, m_{\rm max}) = 1/(m_{\rm max}-m_{\rm min}), ~ \text{for}~ m_{\rm NS} \in{(m_{\rm min}, m_{\rm max})} .
\end{equation}
Motivated by \cite{2010ApJ...725.1918O} and \cite{2021ApJ...913L...7A}, we introduce a Truncated Gaussian and a Truncated Power Law for the mass distribution of BHs,
\begin{equation}\label{TPL}
\pi(m_{\rm BH}|m_{\rm low}, m_{\rm up}, \mu, \sigma) = \mathcal{N}(m_{\rm BH}|\mu,\sigma)/\Phi', ~ \text{for}~ m_{\rm BH} \in{(m_{\rm low}, m_{\rm up})} .
\end{equation}
\begin{equation}
\pi(m_{\rm BH}|m_{\rm low}, m_{\rm up}, \alpha) = \frac{m_{\rm BH}^{-\alpha}}{\int_{m_{\rm low}}^{m_{\rm up}}{m^{-\alpha}dm}}, ~ \text{for}~ m_{\rm BH} \in{(m_{\rm low}, m_{\rm up})} .
\end{equation}
In the above equations, $m_{\rm NS}$ and $m_{\rm BH}$ are the mass of NS and BH, while $\Phi_1$, $\Phi_2$, $\Phi$, and $\Phi'$ are the normalization constants, and other parameters are described in Table \ref{table_priors}. 

The parameterization for the component (BH or NS) spin magnitudes and tilts is similar to the \textsc{Default} model in \cite{2021ApJ...913L...7A}. The spin tilt of each component in a binary is assumed to be independently drawn from the same underlying distribution, see Eq. (\ref{tilt}) below. For simplicity, we use two truncated gaussians to fit the dimensionless spin magnitudes of BHs and NSs (i.e., $a_{\rm BH}$ and $a_{\rm NS}$ ), respectively. Therefore, the model of spin reads
\begin{equation}\label{tilt}
\pi(z| \zeta, \sigma_{\rm t}) = \zeta\mathcal{N}(z|1, \sigma_{\rm t})/\Phi_{\rm z}+(1-\zeta)/2, ~ \text{for}~ z \in{(-1, 1)} ,
\end{equation}
\begin{equation}\label{aBH}
\pi(a_{\rm BH}| \mu_{\rm a}^{\rm BH}, \sigma_{\rm a}^{\rm BH}) = \mathcal{N}(a_{\rm BH}|\mu_{\rm a}^{\rm BH}, \sigma_{\rm a}^{\rm BH})/\Phi_{\rm a}^{\rm BH}, ~ \text{for}~ a_{\rm BH} \in{(0, 1)} ,
\end{equation}
\begin{equation}\label{aNS}
\pi(a_{\rm NS}| \mu_{\rm a}^{\rm NS}, \sigma_{\rm a}^{\rm NS}) = \mathcal{N}(a_{\rm NS}|\mu_{\rm a}^{\rm NS}, \sigma_{\rm a}^{\rm NS})/\Phi_{\rm a}^{\rm NS}, ~ \text{for}~ a_{\rm NS} \in{(0, a_{\rm max}^{\rm NS})} ,
\end{equation}
 where $z = \cos{\theta_{1,2}}$ ($\theta_{1,2}$ is the tilt angle between component spin and binary's orbital angular momentum), $\Phi_{\rm z}$, $\Phi_{\rm a}^{\rm BH}$, and $\Phi_{\rm a}^{\rm NS}$ are the normalization constants. The second term of Eq.  (\ref{tilt}) represents the probability density of systems that have an isotropic spin distribution, and $\zeta$ represents the mixing fraction of the field mergers.
Note that, NSs are able to achieve larger spin tilts since they are lower mass and the orbital plane of the binary can theoretically be tilted a larger degree due to the impact of the supernova \citep{2016ApJ...832L...2R}. For simplicity, we assume the same tilt angle distribution for both BHs and NSs, since the currently limited data are not able to recognize such difference between BHs and NSs.
The highest dimensionless NS spin implied by pulsar-timing observations of binaries that merge within a Hubble time is ∼0.04 \citep{2018ApJ...854L..22S, 2018PhRvD..98d3002Z}, therefore, it's reasonable to set $a_{\rm max}^{\rm NS}=0.05$. All the description of the parameters and the priors of each model are summarized in Table \ref{table_priors}

Combining the mass and spin models, we get the synthesis models for NSs and BHs,
\begin{equation}\label{synthesis_NS}
\pi(m_{\rm NS}, a_{\rm NS}, z| \boldsymbol{\Lambda_{\rm NS}}) = \pi(m_{\rm NS}| \boldsymbol{\Lambda_{\rm NS}^{\rm m}})\pi(a_{\rm NS}| \boldsymbol{\Lambda_{\rm NS}^{\rm a}})\pi(z| \boldsymbol{\Lambda_{\rm z}}) ,
\end{equation}
\begin{equation}\label{synthesis_BH}
\pi(m_{\rm BH}, a_{\rm BH}, z| \boldsymbol{\Lambda_{\rm BH}}) = \pi(m_{\rm BH}| \boldsymbol{\Lambda_{\rm BH}^{\rm m}})\pi(a_{\rm BH}| \boldsymbol{\Lambda_{\rm BH}^{\rm a}})\pi(z| \boldsymbol{\Lambda_{\rm z}}) ,
\end{equation}
where $\boldsymbol{\Lambda_{\rm NS}^{\rm m}}$ ($\boldsymbol{\Lambda_{\rm BH}^{\rm m}}$), $\boldsymbol{\Lambda_{\rm NS}^{\rm a}}$ ($\boldsymbol{\Lambda_{\rm BH}^{\rm a}}$), and $\boldsymbol{\Lambda_{\rm z}}$ are the  population parameters for mass distribution of NS (BH), spin magnitude of NS (BH), and spin orientations, respectively. And $\boldsymbol{\Lambda_{\rm NS}}=\boldsymbol{\Lambda_{\rm NS}^{\rm m}}\cup\boldsymbol{\Lambda_{\rm NS}^{\rm a}}\cup\boldsymbol{\Lambda_{\rm z}}$, $\boldsymbol{\Lambda_{\rm BH}}=\boldsymbol{\Lambda_{\rm BH}^{\rm m}}\cup\boldsymbol{\Lambda_{\rm BH}^{\rm a}}\cup\boldsymbol{\Lambda_{\rm z}}$.
Since the secondary objects of all the binaries are NSs,  the model for them is
\begin{equation}\label{synthesis_2}
\pi(m_{\rm 2}, a_{\rm 2}, z_2| \boldsymbol{\Lambda_{\rm 2}}) = \pi(m_{\rm NS}, a_{\rm NS}, z_2| \boldsymbol{\Lambda_{\rm NS}}) .
\end{equation}
While the primary objects can be NSs or BHs, we define $r_{\rm BNS}$ as the fraction of BNS populations, then the model for primary objects becomes 
\begin{equation}\label{synthesis_1}
\pi(m_{\rm 1}, a_{\rm 1}, z_1| \boldsymbol{\Lambda_{\rm 1}}) = r_{\rm BNS}\pi(m_{\rm NS}, a_{\rm NS}, z_1| \boldsymbol{\Lambda_{\rm NS}}) + (1-r_{\rm BNS})\pi(m_{\rm BH}, a_{\rm BH}, z_1| \boldsymbol{\Lambda_{\rm BH}}).
\end{equation}
So the synthesis models for the NSBH and BNS binaries are represented by
\begin{equation}\label{synthesis}
\pi(\theta|\boldsymbol{\Lambda})=\pi(m_{\rm 2}, a_{\rm 2}, z_2,m_{\rm 1}, a_{\rm 1}, z_1| \boldsymbol{\Lambda}) = \pi(m_{\rm 2}, a_{\rm 2}, z_2| \boldsymbol{\Lambda_{\rm 2}})\pi(m_{\rm 1}, a_{\rm 1}, z_1| \boldsymbol{\Lambda_{\rm 1}}) ,
\end{equation}
where $\theta$ is the source parameters for each event, and $ \boldsymbol{\Lambda} = \boldsymbol{\Lambda_{\rm 1}}\cup\boldsymbol{\Lambda_{\rm 2}} = \boldsymbol{\Lambda_{\rm NS}}\cup\boldsymbol{\Lambda_{\rm BH}} \cup \{r_{\rm BNS}\}$ is the population parameters for synthesis models.
Additionally, in order to test whether the distribution of NSs observed via GW is consistent with that observed electromagnetically in the Galaxy, we fix the parameters of Double Gaussian for NSs as $m_{\rm min}=1M_{\odot},~ m_{\rm max}=2.25M_{\odot}, ~\mu_1=1.36M_{\odot}, ~\sigma_1=0.09M_{\odot}, ~\mu_2=1.9M_{\odot},~ \sigma_2=0.5M_{\odot}, ~{\rm and}~r_1=0.65$ (i.e., the median values of the parameters taken from \cite{2020PhRvD.102f3006S}, where the NS mass distribution was obtained based on the observed systems involving at least one NS in the Galaxy).
Thus, there are 4 NS mass distribution models (i.e., Double Gaussian\footnote{Note that the Double Gaussian with free parameters is not exactly the distribution of Galactic NSs, since the inferred parameters from GW may be inconsistent with that from Galactic NSs}, Single Gaussian, Uniform, and Galactic distribution) $\times$ 2 BH mass distribution models (i.e., Truncated Power Law and Truncated Gaussian) $\times$ 1 spin model. In other words we have  8 synthesis models. 

\begin{table}[htpb]
\begin{ruledtabular}
\caption{Distribution of the priors for hierarchical Bayesian inference.}
\label{table_priors}
\begin{tabular}{ccccc}
Description  &Parameters       & \multicolumn{3}{c}{NS mass distribution priors}     \\ \cline{3-5}
                   && Double Gaussian                   & Single Gaussian & Uniform     \\ \hline
 minimum mass of NSs&$m_{\rm min}[M_{\odot}]$        & U(0.9,1.3)             &  { U(0.9,1.3)}&  { U(0.9,1.3)}  \\
 maximum mass of NSs&$m_{\rm max}[M_{\odot}]$        & U(1.7,2.9)            &  U(1.7,2.9)&  U(1.7,2.9)   \\
 mean of first Gaussian&$\mu_1[M_{\odot}]$     & U(1, 2.2)& - & - \\
 mean of second Gaussian&$\mu_2[M_{\odot}]$             & U($1, 2.2$)          & -  & -    \\
 standard deviation of first Gaussian&$\sigma_1[M_{\odot}]$             & U($0.01,1$)           & -    & -  \\
 standard deviation of second Gaussian&$\sigma_2[M_{\odot}]$             & U($0.01,1$)            & - & - \\
fraction of first Gaussian& $r_1$             & U($0,1$)          & - & -  \\
 mean of single Gaussian&$\mu[M_{\odot}]$     & - & U(1, 2.2) & - \\
standard deviation of single Gaussian& $\sigma[M_{\odot}]$            & -           & U($0.01,1$) & -  \\
 fraction of BNS population&$r_{\rm BNS}$   & U($0,1$)       & U($0,1$)   & U($0,1$)  \\
\multirow{2}{*}{Constraint} & - & $m_{\rm min}<\mu_1<\mu_2<m_{\rm max}$ & $m_{\rm min}<\mu<m_{\rm max}$ & - \\
 &-& $\sigma_1<\sigma_2$ & $ - $ & - \\
 \hline
 & & \multicolumn{3}{c}{BH mass distribution priors}     \\ \cline{3-5}
 & & Truncated Power Law              &    & Truncated Gaussian    \\ \hline
low boundary of BH mass distribution &$m_{\rm low}[M_{\odot}]$        & { U(3,7)}        &    & { U(3,7)}  \\
high boundary of BH mass distribution & $m_{\rm up}[M_{\odot}]$        & { U(8,30)}        &    & { U(8,30)}  \\
negative spectral index of Power law& $\alpha$             & U($-4, 12$)         & & -          \\
mean of Gaussian for BH& $\mu_{\rm BH}[M_{\odot}]$       &  - &        & U($5,15$)            \\
standard deviation of Gaussian for BH&$\sigma_{\rm BH}[M_{\odot}]$    &  -   &           & U($0.5,10$)           \\
Constraint &- & - &  & $m_{\rm low}<\mu_{\rm BH}<m_{\rm up}$ \\
  \hline
& & \multicolumn{3}{c}{spin priors}     \\ \cline{1-5}
mean of  spin for NS & $\mu_{\rm a}^{\rm NS}$             &   & U($0, 0.05$) &      \\
standard deviation of spin for NS& $\sigma_{\rm a}^{\rm NS}$             &            & { U($0.001,0.05$)}  &     \\
mean of spin for BH &  $\mu_{\rm a}^{\rm BH}$        &     & U($0, 0.99$)          &     \\
standard deviation of spin for BH& $\sigma_{\rm a}^{\rm BH}$      &       & U($0.01,0.25$)           &     \\
width of spin misalignment for field mergers& $\sigma_{\rm t}$             &            & U($0.01,4.$)    &    \\
fraction of field mergers & $\zeta$   &        & U($0,1$)     &   \\

\end{tabular}
\tablenotetext{}{{\bf Note.} Here, `U' means the uniform distribution.}
\end{ruledtabular}
\end{table}

\subsection{Hierarchical likelihood}
With the data of the observed events $\{d\}$ that described in Sec. \ref{method_data}, and the population models described in Sec. \ref{method_model}, we perform a hierarchical Bayesian inference following \cite{2021ApJ...913L...7A} and \cite{2019PASA...36...10T}. For the given data $\{d\}$ from $N_{\rm det}$ GW detections, the likelihood of $\boldsymbol{\Lambda}$ can be expressed as 
\begin{equation}\label{eq_llh}
\mathcal{L}(\{d\}|\boldsymbol{\Lambda})\propto\prod_{i=1}^{N_{\rm det}}\frac{\int{\mathcal{L}(d_i|\theta_i)\pi(\theta_i|\boldsymbol{\Lambda})d\theta_i}}{\xi_i(\boldsymbol{\Lambda})},
\end{equation} 
where $\xi_i(\boldsymbol{\Lambda})$ means the detection fraction,
and the single-event likelihood $\mathcal{L}(d_i|\theta_i)$ can be estimated using the posterior samples described in Sec. \ref{method_data} (see \cite{2021ApJ...913L...7A} for detail), then the Eq. (\ref{eq_llh}) takes the form of 
\begin{equation}
\mathcal{L}(\{d\}|\boldsymbol{\Lambda}) \propto \prod_{i=1}^{N_{\rm det}} \frac{1}{\xi_i(\boldsymbol{\Lambda})} \frac{1}{n_i} \sum_{k}^{n_i}{\frac{\pi(\theta_i^k|\boldsymbol{\Lambda})}{\pi_{\phi}(\theta_i^k)}},
\end{equation} 
where $\pi_{\phi}(\theta_i^k)$ is the default prior used for parameter estimation of each individual event, and $n_i$ is the size of posterior samples.
For the estimation of $\xi_i(\boldsymbol{\Lambda})$, we use a Monte Carlo integral over detected injections as introduced in \citet{2021ApJ...913L...7A}  and  \cite{2018CQGra..35n5009T}. And the injection campaigns are made by injecting simulated signal to the LIGO-Virgo detectors with noise curves\footnote{We take the “O3actual” PSD, which is available at \url{https://dcc.ligo.org/LIGO-T2000012/public}, for the events except GW170817. Since GW170817 was detected in O2, the noise curve  can be approximated by the Early High Sensitivity curve in \cite{2018LRR....21....3A}, and can be found at \url{https://git.ligo.org/lscsoft/bilby/-/tree/master/bilby/gw/detector/noise_curves}}. We define the threshold of the injected signal being detected as the network signal-to-noise ratio greater than 12. 
For the hierarchical Bayesian inference, we use the Bilby package \citep{2019ascl.soft01011A} and Dynesty sampler \citep{2020MNRAS.493.3132S}.

\section{Results}\label{sec:result}
In this section we report the results obtained by the hierarchical inference using synthesis models described in Sec. \ref{sec:method}. 
We calculate the Bayes factor for each model ($M_a$) relative to the model of Double Gaussian + Truncated Gaussian ($M_b$) as $\mathcal{B}^{a}_{b} = {Z_{a}}/{Z_{b}}$, where $Z_a$ and $Z_b$ are the evidences for the model a and b. Formally, the correct metric to compare two models is the odds ratio $\mathcal{O}^a_b = (Z_a/Z_b)\cdot(\pi_a/\pi_b)$, we assume the prior odds $\pi_a$ and $\pi_b$ for the models are equal, so that $\mathcal{O}^a_b=\mathcal{B}^a_b$.
With the logarithm Bayes factors summarized in Table \ref{table_BF}, we can conclude that all three NS mass function models are comparable to fit the data. Nevertheless, it is interesting to see that the Galactic model (i.e., the fixed Double Gaussian model) is slightly preferred than the others.  
It is certainly too early to conclude that the mass distribution of NSs with GW is the same as that of the Galactic NSs, and much more data are needed to draw a robust conclusion.
As for BH mass distribution, we find that the Truncated Power Law model has a preference over the Truncated Gaussian by $\ln\mathcal{B}=\sim 0.5 -1.5$.

\begin{table}[htpb]
\begin{ruledtabular}
\caption{Logarithm Bayes factors of the NS and BH mass distribution models.}\label{table_BF}
\centering
\begin{tabular}{ccccc}
    $\ln{\mathcal{B}}$     & \multicolumn{4}{c}{NS mass distribution models}     \\ \cline{2-5}
  BH mass distribution models                 & Double Gaussian                   & Single Gaussian & Uniform  & Galactic   \\ \hline
 Truncated Gaussian        & 0            &  0.21 &  -0.15   &   1.37\\
 Truncated Power Law       & 0.78         &  0.63  &  0.82  & 2.68\\
\end{tabular}
\tablenotetext{}{{\bf Note.} The values are relative to the evidence of Double Gaussian (NS mass distribution model) + Truncated Gaussian (BH mass distribution model).}
\end{ruledtabular}
\end{table}

\begin{figure*}
\centering
\gridline{\fig{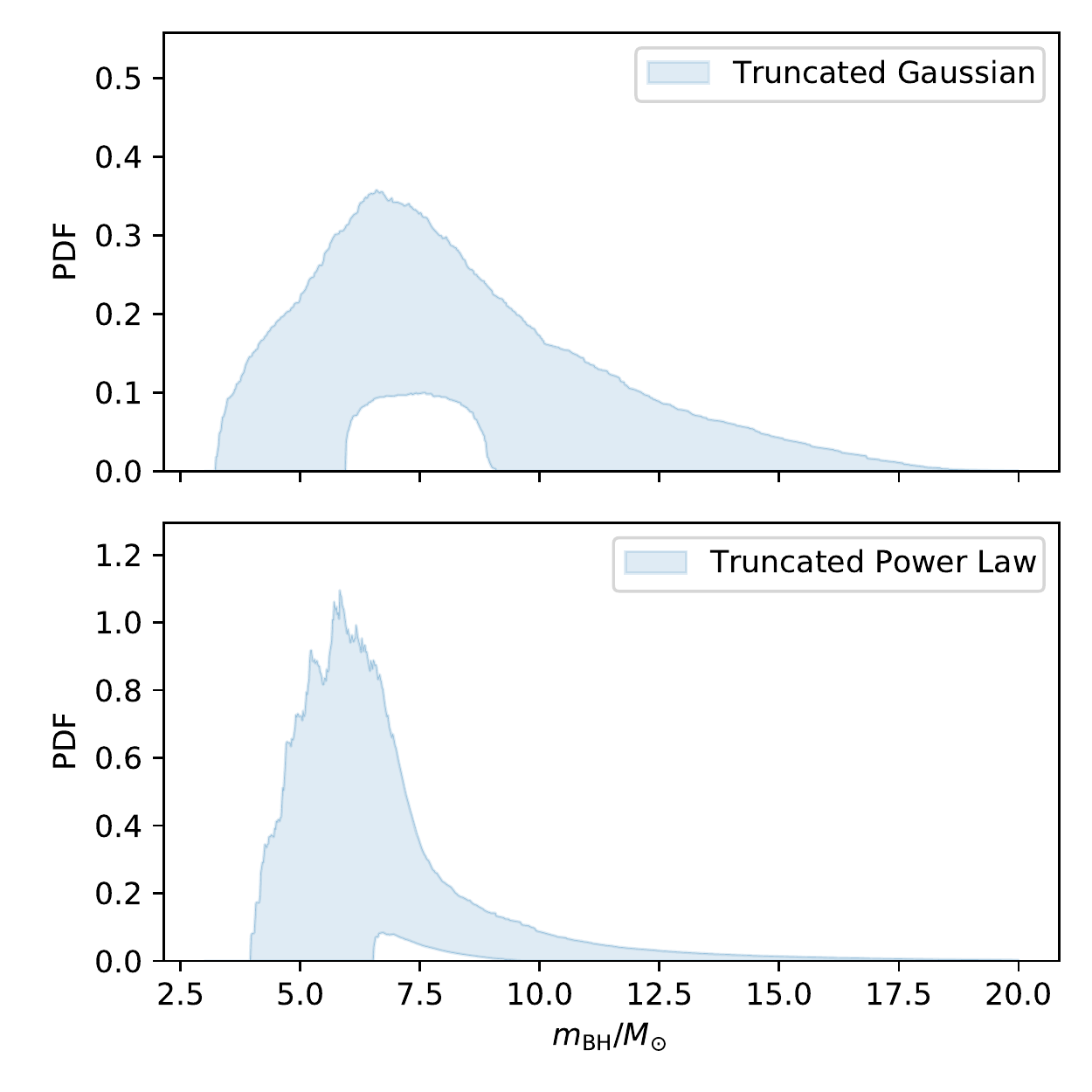}{0.5\textwidth}{}
\fig{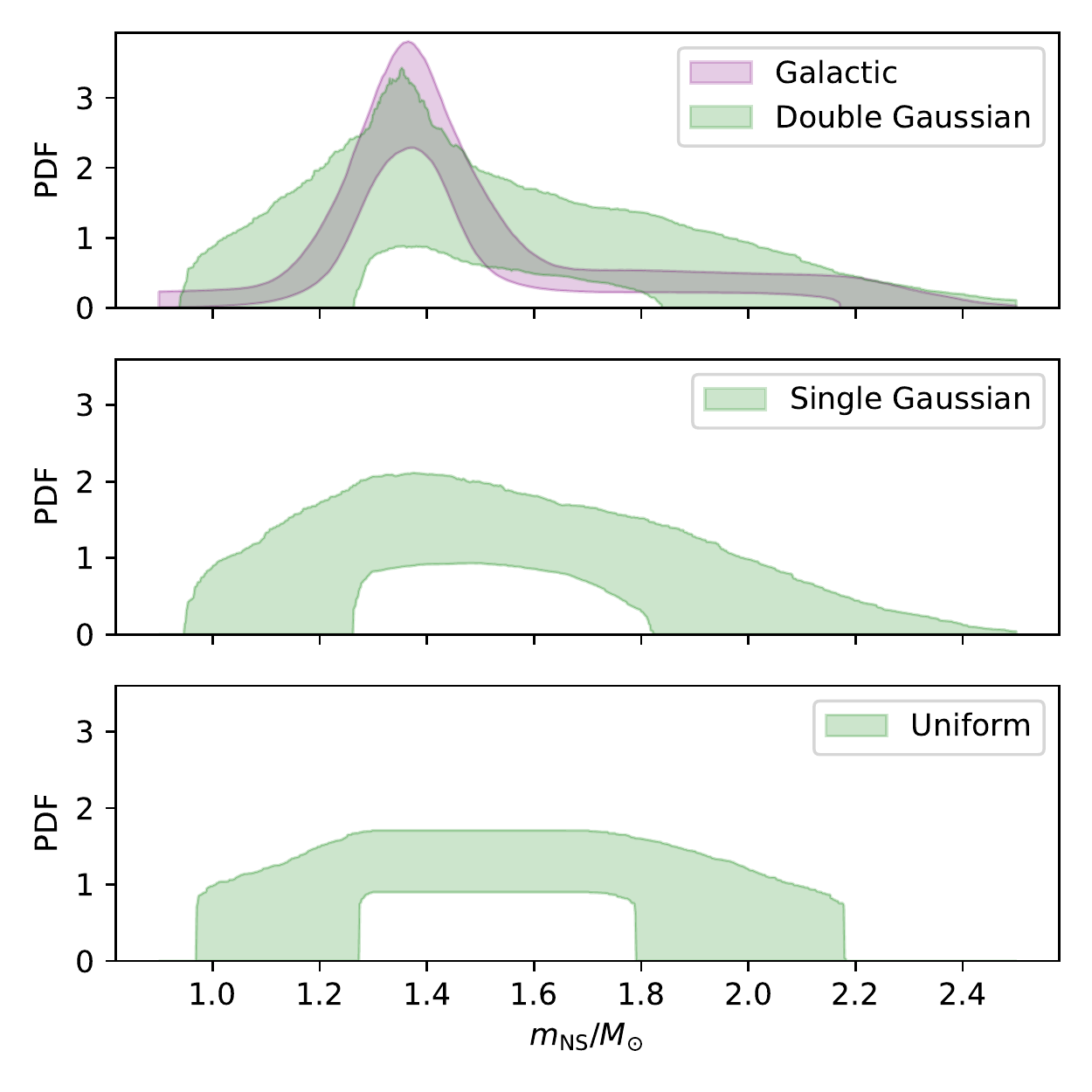}{0.5\textwidth}{}}
\caption{Mass distributions of NSs and BHs. The left column are the BH mass distributions obtained by Truncated Gaussian and Truncated Power Law, assuming the NS mass distribution model of Double Gaussian. The right column are the NS mass distributions obtained by Double Gaussian, Single Gaussian, and Uniform models, respectively, and their partner model for BH mass is the Truncated Power Law. 
The purple shaded region represents the mass distribution of the Galactic NSs as obtained by \cite{2020PhRvD.102f3006S}.
In each panel, the shaded region represents the 90\% credible interval. }
\label{massdist}
\end{figure*}

The inferred population parameters for NS and BH mass distribution models are presented in Table \ref{table_NS}. The results for NS mass distribution are inferred  by the Double Gaussian, Single Gaussian, and Uniform models, assuming a BH mass distribution model of Truncated Power Law, and the results for BH mass distribution are obtained using Truncated Power Law and Truncated Gaussian models, assuming a NS mass distribution model of Double Gaussian. Actually, the results assuming other BH mass distribution model/ NS mass distribution models are rather similar. In order to build more intuition, we plot the mass distributions of BHs/NSs as shown in Fig. \ref{massdist}. For the Double Gaussian model of NS mass distribution, it is clear that there is a peak around $1.35M_{\odot}$, which is consistent with the first peak found in the mass distribution of Galactic NSs, but there is no  peak in the higher mass range (around $1.9M_{\odot}$, for example). 
As for the Single Gaussian model, there is no significant peak. In comparison to the Uniform model, there is a moderate bulge in the low mass range (i.e., $\sim 1.2M_{\odot} - 1.6M_{\odot}$). 
The low mass cutoff of the BHs is constrained to $5.55^{+0.96}_{-1.52}M_{\odot}$ ($4.61^{+1.32}_{-1.37}M_{\odot}$) by the Truncated Power Law (Truncated Gaussian) model, however the high mass cutoff of BHs in NSBHs is constrained poorly by either model (i.e., $17.41^{+11.00}_{-7.89}M_{\odot}$ and $14.26^{+13.59}_{-5.21}M_{\odot}$ for Truncated Power Law and Truncated Gaussian models). The spectrum index of Truncated Power Law is $5.95^{+4.70}_{-4.97}$, which is consistent with the constraints of the \cite{2021ApJ...913L...7A}, while the posterior support pushes to higher values. 
Interestingly, the mass distribution of BHs obtained by the Truncated Gaussian, i.e., $\mu_{\rm BH}=6.97^{+2.92}_{-1.66}M_{\odot}$ and $\sigma_{\rm BH}=3.47^{+5.39}_{-2.30}M_{\odot}$ is consistent with that constrained by the observations of X-ray binaries in the Galaxy \citep{2010ApJ...725.1918O}.
The fraction of BNS mergers in the NSBH and BNS population is $0.68^{+0.20}_{-0.31}$ (assuming Truncated Power Law + Double Gaussian model), and it is rather similar in all the synthesis models. This estimated fraction $r_{\rm BNS}$ is consistent with the merger rate densities of BNS (i.e., $320^{+490}_{-240}\rm{Gpc^{-3}yr^{-1}}$ obtained by \cite{2021PhRvX..11b1053A}) and NSBH (i.e., $130^{+142}_{-69}\rm{Gpc^{-3}yr^{-1}}$ obtained by \cite{2021ApJ...915L...5A}; see also \cite{2017ApJ...844L..22L} for a NSBH merger rate $\geq 100~{\rm Gpc^{-3}yr^{-1}}$ based on the kilonova modeling).

\begin{table*}[htpb]
\begin{ruledtabular}
\caption{Inferred population parameters of the NS and BH mass distribution models. }

\label{table_NS}
\begin{tabular}{ccccc}
  &Parameters       & \multicolumn{3}{c}{NS mass distribution models assuming Truncated Power Law BH mass distribution model}     \\ \cline{3-5}
                   && Double Gaussian                   & Single Gaussian & Uniform     \\ \hline
&$m_{\rm min}[M_{\odot}]$        &  $1.12^{+0.15}_{-0.18}$           &  $1.12^{+0.14}_{-0.18}$ &  $1.16^{+0.12}_{-0.19}$ \\
&$m_{\rm max}[M_{\odot}]$        & $2.13^{+0.67}_{-0.29}$            &  $2.10^{+0.71}_{-0.28}$ & $1.94^{+0.25}_{-0.16}$   \\
&$\mu_1$($\mu$)$[M_{\odot}]$     & $1.35^{+0.25}_{-0.22}$ & $ 1.42^{+0.36}_{-0.26}$ & - \\
&$\mu_2[M_{\odot}]$            & $1.68^{+0.36}_{-0.33}$       & -  & -    \\
&$\sigma_1$($\sigma$)$[M_{\odot}]$             & $0.28^{+0.40}_{-0.23}$           & $0.47^{+0.42}_{-0.25}$    & -  \\
&$\sigma_2[M_{\odot}]$             & $0.67^{+0.29}_{-0.39}$           & - & - \\
& $r_1$             & $0.58^{+0.37}_{-0.46}$ & - & -  \\
&$r_{\rm BNS}$   & $0.68^{+0.20}_{-0.31}$       & $0.67^{+0.20}_{-0.30}$   & $0.67^{+0.21}_{-0.31}$  \\
 \hline
 & & \multicolumn{3}{c}{BH mass distribution models assuming NS mass distribution model of Double Gaussian}     \\ \cline{3-5}
 & & Truncated Power Law              &    & Truncated Gaussian    \\ \hline
&$m_{\rm low}[M_{\odot}]$        & $5.55^{+0.96}_{-1.52}$       &    & $4.61^{+1.32}_{-1.37}$  \\
& $m_{\rm up}[M_{\odot}]$        & $17.41^{+11.00}_{-7.89}$       &    & $14.26^{+13.59}_{-5.21}$  \\
& $\alpha$             & $5.95^{+4.70}_{-4.97}$         & & -          \\
& $\mu_{\rm BH}[M_{\odot}]$       &  - &        & $ 6.97^{+2.92}_{-1.66}$            \\
&$\sigma_{\rm BH}[M_{\odot}]$    &  -   &           & $3.47^{+5.39}_{-2.30}$           \\
\end{tabular}
\tablenotetext{}{{\bf Note.} The values represent the medians and symmetric 90\% credible intervals of the parameters.}
\end{ruledtabular}
\end{table*}

\begin{figure*}
\centering
\includegraphics[scale=0.4]{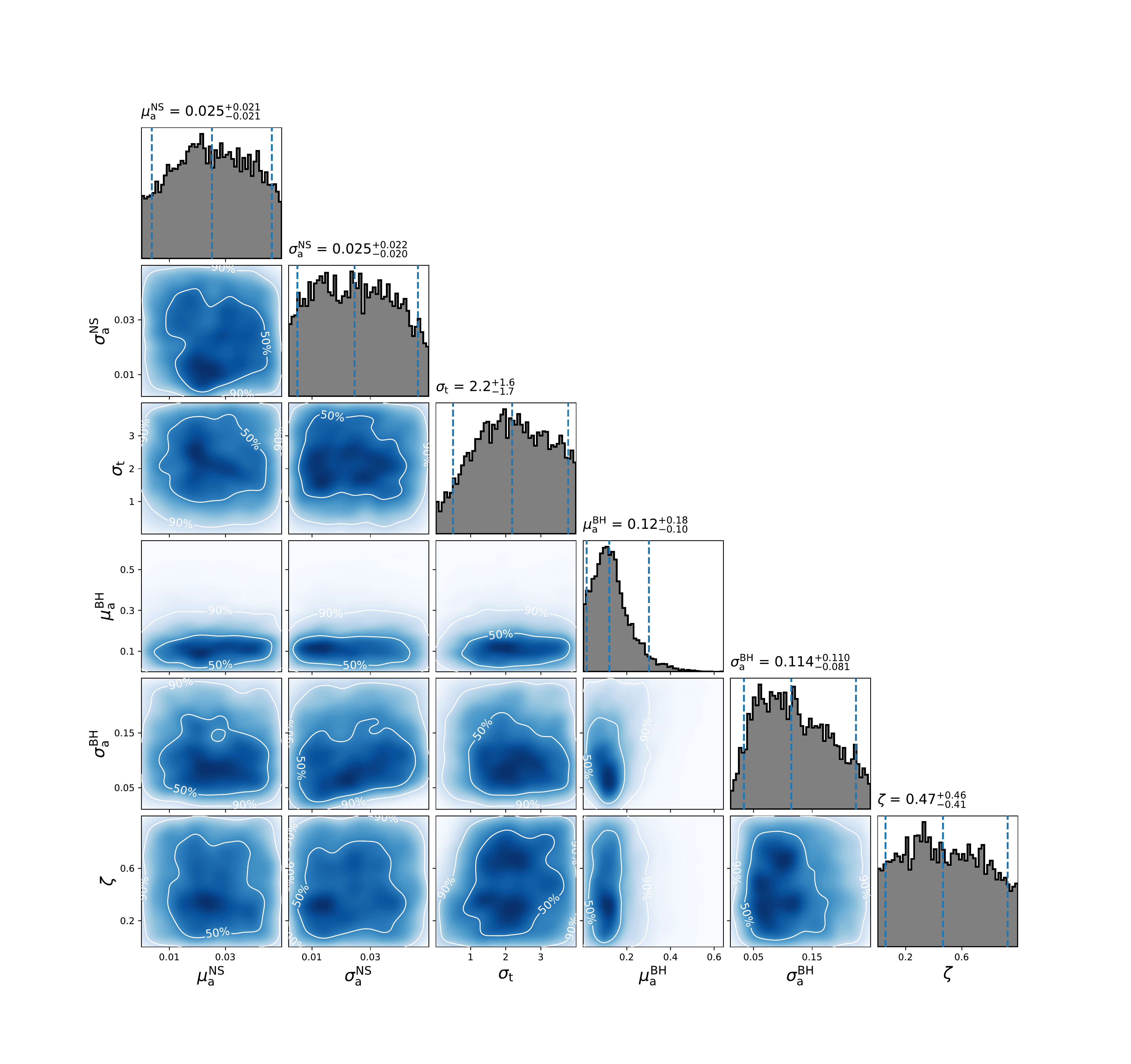}
\caption{Posterior distribution for the spin model described in Sec. \ref{method_model}, assuming the Truncated Power Law BH mass function model and Double Gaussian NS mass distribution model. The contours represent 50\% and 90\% credible bounds, respectively.}
\label{fig_spin}
\end{figure*}

As for the spin properties of the binaries, the posterior distribution obtained by the Truncated Power Law + the Double Gaussian model is shown in Fig. \ref{fig_spin}, and the results from other models are similar. Though the distribution of misalignment is not well constrained, it is quite clear that a perfectly aligned spin distribution ($\sigma_t = 0$, $\zeta = 1$) has been ruled out (see also \cite{2021ApJ...913L...7A} for the same conclusion for BBHs), while a purely isotropic distribution of spin orientation ($\sigma_t > 2$) is still allowed by the current data.
The distribution of BHs' dimensionless spin magnitudes is constrained to $\mu_{\rm a}^{\rm BH}=0.12^{+0.18}_{-0.10}$ and $\sigma_{\rm a}^{\rm BH}=0.11^{+0.11}_{-0.08}$, but the distribution for NSs is poorly constrained.
We do not expect to determine the population properties of spins of these kinds of binaries with currently limited observations, while the correlation between the spin parameters and the component masses of the compact binaries will influence the inference of the NS mass distribution with the GW data \citep{2013PhRvD..87b4035B,2018JCAP...03..007C,2020PhRvR...2d3096P}, so it is helpful to incorporate the spin information into our population inference.

\section{simulation}\label{sim}
We expect to have dozens of BNS/NSBH detections by the end of the fourth observing run of LIGO/Virgo/KAGRA network, and the number of events may reach one hundred within the duration of their fifth observing run \citep{2018LRR....21....3A}. Thus it is interesting to investigate whether we can determine the mass distribution of the NS via GWs in the next few years. In this section, we perform our analysis on mock GW detections generated from a prespecified underlying population. We assume the underlying NS mass distribution as that of the Galactic distribution, i.e., the Double Gaussian model (see Eq. (\ref{DG})) with $m_{\rm min}=1M_{\odot}$, $m_{\rm max}=2.25$, $\mu_1=1.36M_{\odot}$, $\sigma_1=0.09M_{\odot}$, $\mu_2=1.9M_{\odot}$, $\sigma_2=0.5M_{\odot}$, and $r=0.65$. The BH mass distribution is described by the Truncated Power Law model (see Eq. (\ref{TPL})) with $m_{\rm low}=5M_{\odot}$, $m_{\rm up}=10M{\odot}$, and $\alpha=3$. The underlying spin distribution of binaries is described by Eq. (\ref{tilt}), Eq. (\ref{aBH}), and Eq. (\ref{aNS}), with $\zeta=0.8$, $\sigma_{\rm t}=0.2$, $\mu_{\rm a}^{\rm BH}=0.1$, $\sigma_{\rm a}^{\rm BH}=0.1$, $\mu_{\rm a}^{\rm NS}=0.005$, and $\sigma_{\rm a}^{\rm NS}=0.01$. Additionally, the fraction of BNSs in the surveyed population (including both BNSs and NSBHs) is assumed to 0.7.
In generating mock detections, we assume that the underlying population follows a uniform in comoving volume and source frame time merger rate, with isotropic sky positions and inclinations. The true values of ($m_1$, $m_2$, $a_1$, $a_2$, $z_1$, $z_2$) are drawn from the given distribution as described above. Additionally, $\phi_{\rm 12}$ and $\phi_{\rm jl}$ (i.e., azimuthal angle between the spins of the two components, azimuthal angle between the total binary angular momentum and the orbital angular momentum)  are set to be uniform in (0,~$2\pi$). The \textit{IMRPhenomXPHM} waveform \citep{2018PhRvL.120p1102L,2017PhRvD..95j4004C,2019PhRvD.100b4059K,2021CQGra..38a5006G} is used for generating simulated signals, and the identification of a GW signal is based on a network signal-to-noise ratio threshold  $\rho_{th} = 12$ and the design sensitivity noise curves\footnote{\url{https://dcc.ligo.org/LIGO-T2000012/public}} \citep{2018LRR....21....3A}. 

To obtain the posteriors for each `detected' event, we apply the user-friendly software \textit{Bilby} \citep{2019ascl.soft01011A} with sampler \textit{Pymultinest} \citep{2016ascl.soft06005B}
for parameter estimation. The sampling priors for single event parameter estimation are set as following: detector frame component masses are uniform in $(0.9, 30)M_{\odot}$, spin magnitudes are uniform in (0,~1)/(0,~0.05) for BHs/NSs, the spin orientations are isotropic, the priors of distance and inclination angle are set assuming the binaries are uniform in comoving volume, source frame time, and with isotropic directions. In order to accelerate the inference, we fix the other extrinsic parameters (i.e., the sky coordinate, the polarization angle, the coalescing phase, and the coalescing time) as their injected values.

\begin{figure*}
\centering
\includegraphics[scale=0.3]{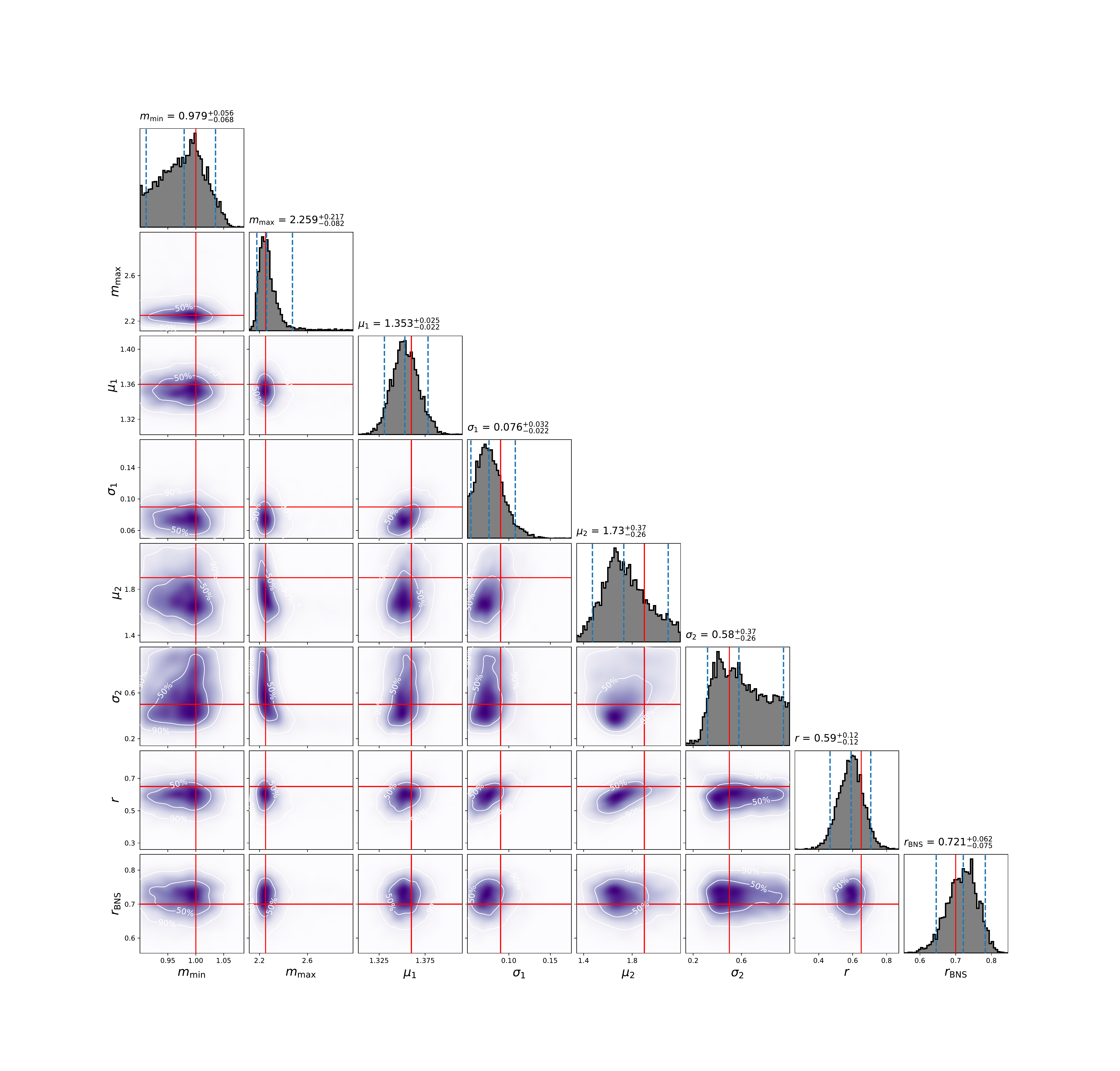}
\caption{Posterior distribution for the NS mass hyperparameters inferred from 100 mock events. The contours represent 50\% and 90\% credible bounds, respectively, and the solid lines stand for the injected hyperparameters.}
\label{NS_sim}
\end{figure*}

\begin{figure*}
\centering
\gridline{\fig{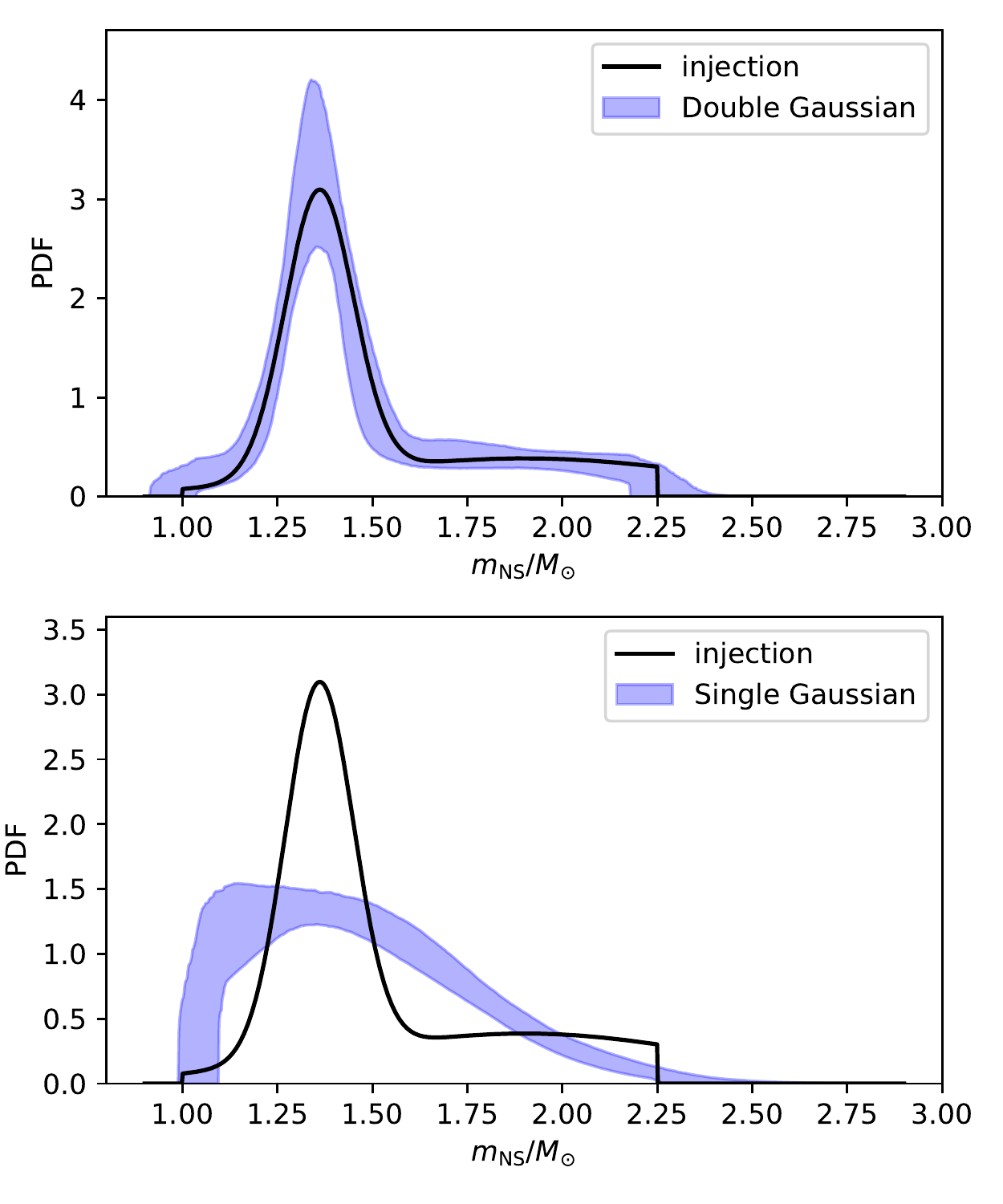}{0.4\textwidth}{}
\fig{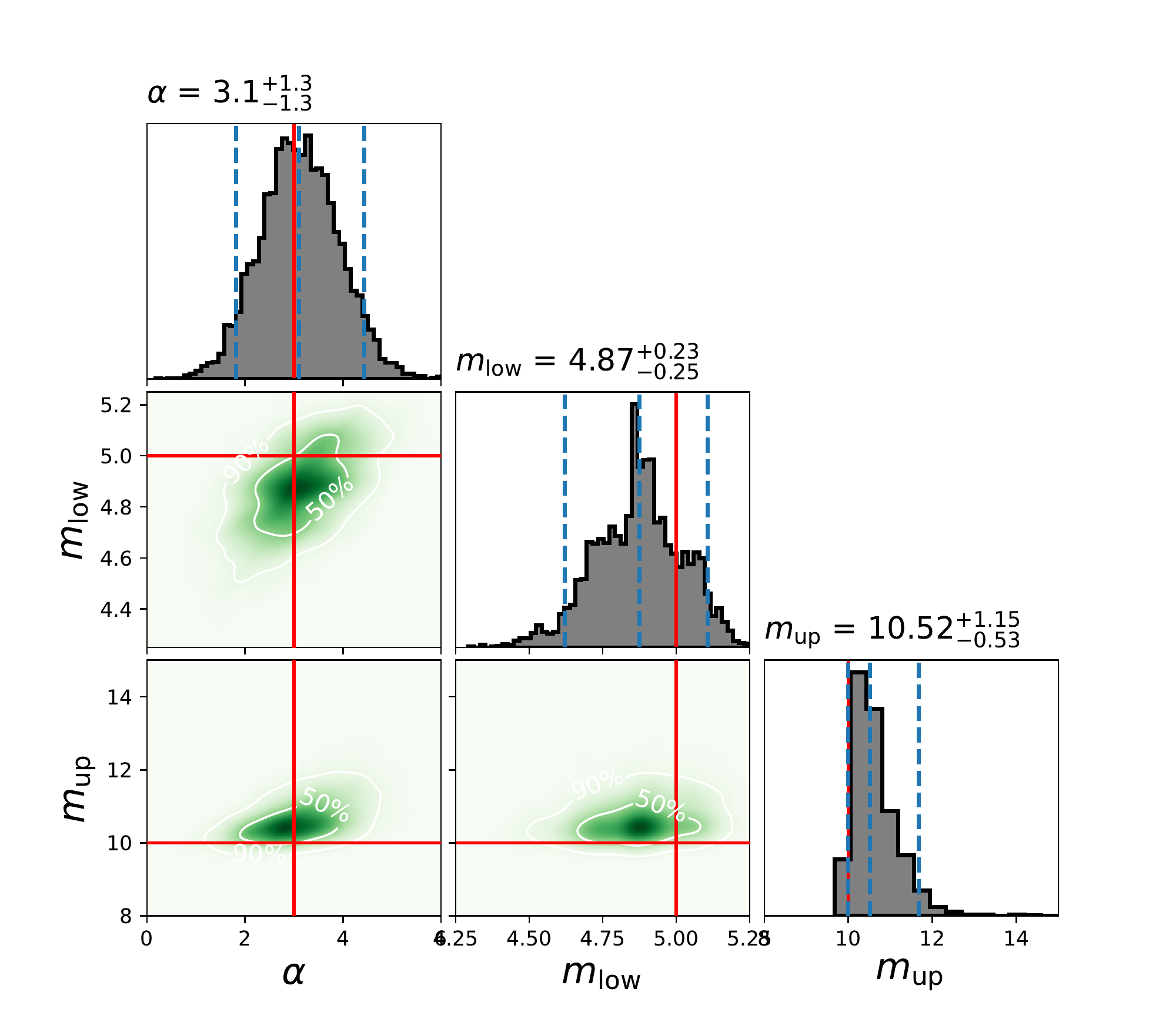}{0.55\textwidth}{}}
\caption{Left: Mass distributions of NSs inferred from the mock GW data with the Double Gaussian and Single Gaussian models. The black curves are the mass distribution of the injected population, and  the shaded region represents the 90\% credible interval. Right: Posterior distribution for the BH hyperparameter parameters inferred from the mock GW data. The contours represent 50\% and 90\% credible bounds, respectively, and the solid lines stand for the injected hyperparameter values.}
\label{NS_BH_sim}
\end{figure*}

We find with 100 `detected' events generated from the simulated population, we can rule out the Uniform distribution for NS mass by $\ln{\mathcal{B}}\sim 39$ (compared with the Double Gaussian model), and the Double gaussian model is more favored than the Single Gaussian model by $\ln{\mathcal{B}}\sim 19$ . The posteriors of hyperparameters for the Double Gaussian model are presented in Fig. \ref{NS_sim}, it is clear that the first peak of the NS mass distribution is well determined, while the second peak is ambiguous. The minimum and maximum masses of NS are both well constrained, additionally, the BNS fraction is constrained within uncertainty of 0.13 at 90\% credible level. 
The distributions of NS mass modeled by the Double Gaussian and the Single Gaussian are shown in Fig. \ref{NS_BH_sim} (left). We can see that for the Double Gaussian model, the first peak is fairly significant, 
while the second peak is not obvious and it looks like a tail of the first peak. As for the Single Gaussian model, it fails to figure out the features of the injected population as is indicated by the Bayes factor mentioned above.
The BH mass distribution is reconstructed by the Truncated Power Law, and all the hyperparameters are constrained well as presented in Fig. \ref{NS_BH_sim} (right). 
However, the spin properties of binaries are not well constrained, where $\mu^{\rm NS}_a=0.02^{+0.003}_{-0.004}$, $\sigma^{\rm NS}_a=0.004^{+0.003}_{-0.002}$,  $\mu^{\rm BH}_a=0.168^{+0.019}_{-0.022}$, $\sigma^{\rm BH}_a=0.071^{+0.024}_{-0.016}$, $\sigma_t=0.87^{+0.20}_{-0.14}$, and $\zeta=0.93^{+0.06}_{-0.19}$.

\section{Conclusion and Discussion}\label{sec:discussion}
We perform hierarchical population inferences for the GW binaries involving at least one NS, and construct several synthesis models that include both the component masses and spin properties of the BNS and NSBH binaries, since there is a correlation between the component masses and the spin properties of compact binaries \citep{2013PhRvD..87b4035B,2020PhRvR...2d3096P,2021ApJ...915L...5A}. We have also performed inferences without spin information for comparison, and find that the corresponding evidence is slightly smaller than the inferences with spin information, e.g., $\ln{\mathcal{B}}=\ln{Z}_{\rm with~spin}-\ln{Z}_{\rm no~spin}=0.5$ for Double Gaussian NS mass distribution model + Truncated Power Law BH mass distribution model \footnote{For the case without spin information, we define the spin model as the default prior $\pi_{\phi}(\theta)$ that is used for single-event parameter estimation.}. The  results are similar to those reported in Table \ref{table_BF}, but the peak of Double Gaussian model in the spin-uninformative case is not as significant as that in the spin-informative case. Therefore, it is worthwhile to take the spin information into consideration when inferring the mass distribution of the compact binaries, though we do not aim to determine the spin properties of these kinds of binaries by currently limited observations. 

Our main conclusion is that the three NS mass distribution models (i.e., Double Gaussian, Single Gaussian, and Uniform) can not be reliably distinguished by the current GW data. Our result  is similar to \cite{2021arXiv210704559L}, where it is found that the Flat (in this work we call the Uniform) and Bimodal (i.e.,  the Double Gaussian in this work) models are equally preferred using the Akaike information criterion \citep{AKAIKE19813}. Nevertheless, there is a significant peak in the mass distribution obtained in our Double Gaussian model, and the location of the peak resembles that of Galactic NSs \citep{2020PhRvD.102f3006S}. We further find that the mass distribution of Galactic NSs \citep{2020PhRvD.102f3006S} is slightly more preferred than the other models (by $\ln\mathcal{B}\sim1.9$), which indicates that the NSs observed via GW signals may have the similar mass distribution as the NSs identified by electromagnetic observation in the Galaxy. This conclusion is different from \cite{2021arXiv210704559L}, where the authors find that the mass distribution of the NSs observed via GWs is unimodal, predicting far fewer low-mass and moderately more high-mass NSs in the population. This difference may be caused by the inclusion of the spin information of compact binaries into our hierarchical population inference. Another reason is that \cite{2021arXiv210704559L} use a uniform distribution for BH masses with a pairing function $p(q)\propto q^{\beta_q}$ for NSBHs, while we use the models for BH masses as indicated by \cite{2010ApJ...725.1918O}. 

Both the BH mass distribution models indicate a narrow distribution located in $\sim 5 - 10M_{\odot}$, which is consistent with the BHs in the Galactic X-ray binaries \citep{2010ApJ...725.1918O}. As for the population of the compact binaries  involving at least one NS, a fraction of $\sim 70\%$ is BNSs, and this result agrees with the merger rate densities of BNS and NSBH estimated by \cite{2021PhRvX..11b1053A} and \cite{2021ApJ...915L...5A}.
Due to the currently limited sample and the measurement uncertainties of NS spins, the distribution of NS spin magnitudes is not well constrained. From the inferred result of spin orientations, we conclude that a perfectly aligned spin distribution can be ruled out, but a purely isotropic distribution of spin orientation is still allowed by the GW data. 

Encouragingly, LIGO/Virgo/KAGRA will start their O4 and O5 observing runs in the next few years, at that time the sensitivities of GW detectors will be significantly improved \citep{2018LRR....21....3A}, and the detection number of such binaries like NSBHs and BNSs will exceed one dozen in the duration of O4 and may reach one hundred by the end of O5. By performing simulations, we find that the mass distribution of NS can be well determined with a total number of 100 events (including NSBHs and BNSs) via GW observations. We do not characterize the mass ratios of the binaries in our models currently, which may be a potential improvement in determining population properties of NSs. For instance, very recently, \cite{2021arXiv210708811T} find that there is a mild evidence for a mass correlation among the two components of the low mass ratio binaries. Therefore, if such a feature exists, the population properties of NSs in the coalescing compact binaries can be characterized more accurately. 
With the observation of NSs in binary systems via GW, we can study the properties of NSs in their final moments, while the radio pulsar observations of Galactic NSs enable us to study the properties of NSs in their midlife. Combining measurements from GW  and radio, provides us with a more complete understanding of compact binaries (involving at least one NS) from formation to merger \citep{2021ApJ...909L..19G}. Additionally, the population difference between the extragalactic NSs and the galactic NSs may have an impact on the determination of the origin of the heavy elements \citep{2017ApJ...851L..18W,2021arXiv210702714C}.
What's more, the reliable measurement of NS mass distribution, enables us to better understand the equation of state of dense matter \citep{2020PhRvD.102f4063C,2020arXiv200101747W}.

 
\acknowledgments

This work was supported in part by NSFC under grants of No. 11921003,  No. 11703098, and and 12073080, the Chinese Academy of Sciences via the Strategic Priority Research Program (Grant No. XDB23040000), Key Research Program of Frontier Sciences (No. QYZDJ-SSW-SYS024). This research has made use of data and software obtained from the Gravitational Wave Open Science Center (\url{https://www.gw-openscience.org}), a service of LIGO Laboratory, the LIGO Scientific Collaboration and the Virgo Collaboration. LIGO is funded by the U.S. National Science Foundation. Virgo is funded by the French Centre National de Recherche Scientifique (CNRS), the Italian Istituto Nazionale della Fisica Nucleare (INFN) and the Dutch Nikhef, with contributions by Polish and Hungarian institutes.\\

\vspace{5mm}
\software{Bilby \citep[version 1.1.4, ascl:1901.011, \url{https://git.ligo.org/lscsoft/bilby/}]{2019ascl.soft01011A},
          Dynesty \citep[version 1.0.1, \url{https://github.com/joshspeagle/dynesty}]{2020MNRAS.493.3132S},
          PyMultiNest \citep[version 2.11, ascl:1606.005, \url{https://github.com/JohannesBuchner/PyMultiNest}]{2016ascl.soft06005B},
          PyCBC \citep[gwastro/pycbc: PyCBC Release v1.16.14, \url{https://github.com/gwastro/pycbc}]{2019PASP..131b4503B,alex_nitz_2021_5347736},
          }

\bibliographystyle{aasjournal}
\bibliography{export-bibtex}

\end{CJK*}
\end{document}